\documentclass{WileyMSP-template}
\usepackage[usenames,dvipsnames]{xcolor}
\usepackage{todonotes}
\usepackage{placeins}
\usepackage{hyperref}
\usepackage{physics}

\begin{document}


\title{In Situ Optimization of an Optoelectronic Reservoir Computer with Digital Delayed Feedback}

\maketitle


\author{Fyodor Morozko$^1$}
\author{Shadad Watad$^1$}
\author{Amir Naser$^1$}
\author{Andrey Novitsky}
\author{Alina Karabchevsky$^{1,3}$*}


\begin{affiliations}

F. Morozko, S. Watad, A. Naser, A. Novitsky, Karabchevsky\\
$^1$School of Electrical and Computer Engineering\\
Ben-Gurion University of the Negev\\
Beer-Sheva 8410501, Israel\\
E-mail: alinak@bgu.ac.il\\

A. Novitsky\\
$^2$ Belarusian State University\\
Minsk, 220030, Belarus\\

A. Karabchevsky\\
$^3$Department of Physics\\
Lancaster University\\
LA1 4YB, United Kingdom

\end{affiliations}


\keywords{reservoir computing, neuromorphic computing, physical computing, optoelectronic oscillator, in situ optimization}\\


\begin{abstract}
\textbf{Abstract}\\
\justifying

Reservoir computing (RC) is a powerful computational framework that addresses the need for efficient, low-power, and high-speed processing of time-dependent data. While RC has demonstrated strong capabilities in signal processing and pattern recognition, its practical deployment in physical hardware is hindered by a critical challenge: the lack of efficient, scalable parameter optimization methods for real-world implementations. Traditionally, RC optimization has relied on software-based modeling, which limits the adaptability and efficiency of hardware-based systems, particularly in high-speed and energy-efficient computing applications. Herein, an in situ optimization approach was employed to demonstrate an optoelectronic delay-based RC system with digital delayed feedback, enabling direct, real-time tuning of system parameters without reliance on external computational resources. By simultaneously optimizing five parameters, normalized mean squared error (NMSE) of 0.028, 0.561, and 0.271 is achieved in three benchmark tasks: waveform classification, time series prediction, and speech recognition outperforming simulation-based optimization with NMSE 0.054, 0.543, and 0.329, respectively, in the two of the three tasks.  This method enhances the feasibility of physical reservoir computing by bridging the gap between theoretical models and practical hardware implementation.
\end{abstract}

\section{Introduction}
Reservoir computing (RC) has emerged as a powerful neuromorphic computing framework for processing time-dependent data, offering advantages in speed, energy efficiency, and ease of training compared to conventional machine learning approaches. By leveraging a fixed, high-dimensional nonlinear transformation of input signals, RC circumvents the need for extensive weight optimization, making it particularly attractive for hardware implementations in optical, electronic, and mechanical systems. However, despite its potential, a key limitation remains: the lack of efficient, scalable optimization strategies for tuning system parameters in physical RC implementations. Traditional approaches rely heavily on software-based modeling, which is impractical for real-time, adaptive hardware systems. Addressing this challenge is crucial to unlocking the full potential of RC for applications in ultrafast signal processing, real-time decision-making, and low-power embedded AI.\\

\begin{figure}[h]
  \centering
  \includegraphics[width=0.8\linewidth]{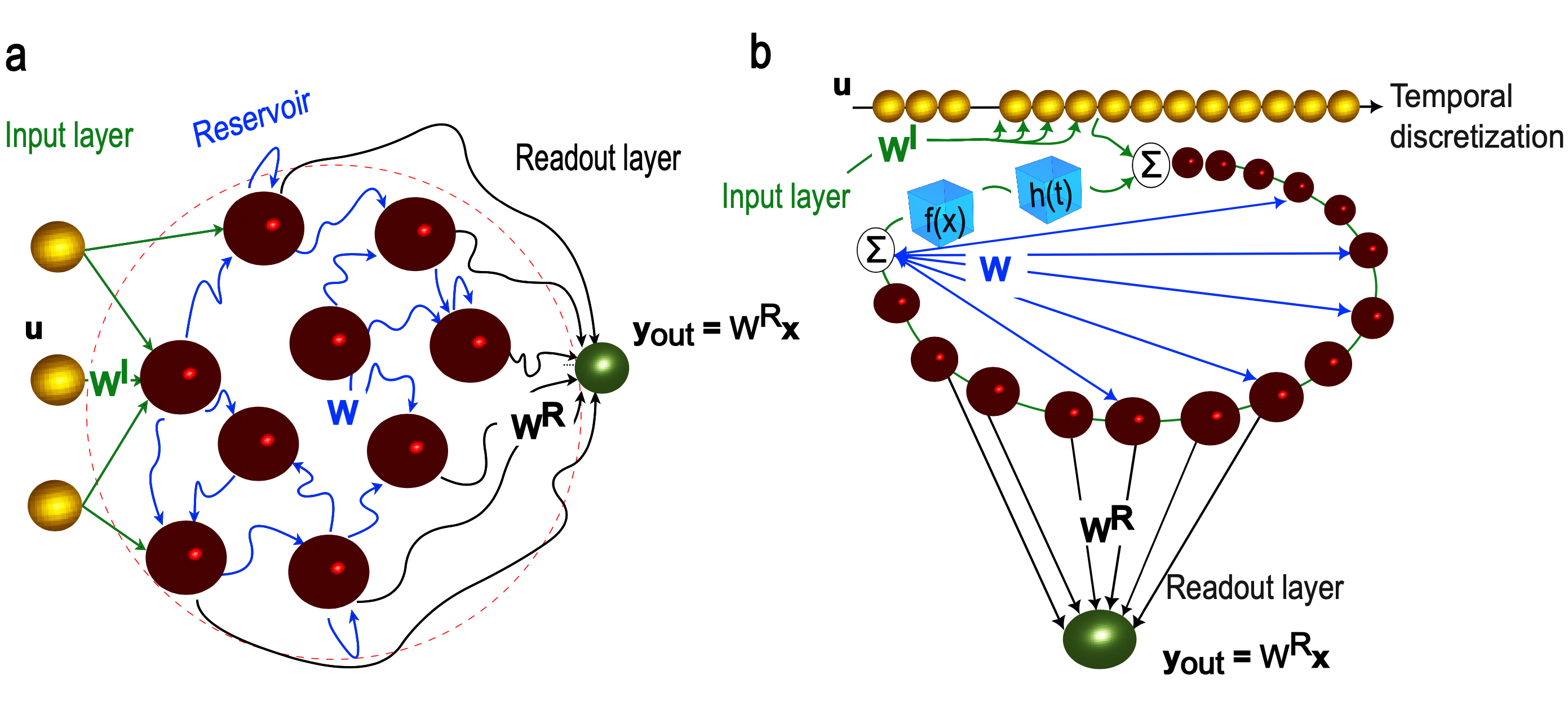}
10  \caption{\textbf{a} Generic reservoir computing principle. The depicted layout consists of distinct components: an input layer (bronze spheres) responsible for receiving external data, a reservoir (ruby spheres) featuring randomized fixed connections, and a linear readout layer (green spheres). \textbf{b} Delay reservoir computing. $f(x)$ is the activation function performing the nonlinear transformation exhibited by the element, and $h(t)$ is the impulse response. $W$, $W^I$ and $W^R$ are, respectively, reservoir, input, and readout connectivity matrices.}
  \label{fig:rc_scheme}
\end{figure}



Traditional RC optimization relies on software simulations, which may not accurately capture the complexities of physical systems, such as noise, non-idealities, and hardware variability~\cite{ref:appeltant2011}~\cite{ref:paquot2012,ref:larger2012,ref:duport2012,ref:martinenghi2012,ref:brunner2013,ref:antonik2017,ref:harkhoe2020}. In situ optimization directly accounts for these real-world factors, leading to more robust and reliable performance. Physical RC systems, especially optoelectronic ones, operate in dynamic environments where temperature fluctuations, component aging, and external disturbances can affect performance. An in situ approach, however, can enable real-time parameter tuning, allowing the system to adapt and maintain optimal functionality without requiring offline recalibration, enhancing the computational efficiency. Software-based optimization often involves time-consuming simulations and iterative tuning, which may not be feasible for high-speed, real-time applications. In situ optimization allows direct tuning within the hardware, reducing computational overhead and enabling faster convergence to optimal operating conditions. By enabling direct, real-time parameter tuning, in situ optimization may transform optoelectronic delay-based RC from a theoretical concept into a practical, high-performance computing tool with broad applications in AI, signal processing, and beyond.

Herein, we address the challenge of optimizing an optoelectronic delay-based reservoir computing (RC) system through an \emph{in situ} approach. We optimize five key hyperparameters simultaneously: the delay-to-clock cycle ratio, input scaling, phase bias, nonlinearity gain, and the regularization parameter used in readout training. The effectiveness of this \emph{in situ} optimization is demonstrated across three benchmark tasks: classification of sinusoidal versus rectangular waveforms, NARMA10 time-series prediction, and Japanese vowel classification. Our method achieves state-of-the-art normalized mean squared error (NMSE) values of 0.028, 0.561, and 0.271, respectively.

Furthermore, we experimentally validate theoretical predictions from Refs. ~\cite{ref:koster2021,ref:hulser2022} which were never realized, and show that resonances between the delay and clock cycle timescales adversely affect RC accuracy, particularly in the NARMA10 task. This is evidenced by significant variations in NMSE, ranging from 0.561 to over 1, depending on the delay-to-clock cycle ratio. These findings provide valuable insights into optimizing the performance of delay-based RC systems for various computational tasks.

\section{Results}
\subsection{Device performance}
We implemented a delay system in an optoelectronic oscillator, a system consisting of a laser light source, an electro-optic modulator, a photodetector whose output is connected to the modulation port of the modulator forming a closed loop. 
Sinusoidal transmission characteristic of a Mach-Zehnder electro-optic modulator renders the system nonlinear while introducing a delay into the feedback loop leads to the emergence of multiple-valued stationary states, causing the system to exhibit complex nonlinear behavior~\cite{ref:ikeda1979,ref:ikeda1980}.
Particularly, optoelectronic oscillator, upon the tuning of laser power, photodetector gain, and phase bias of the modulator, can exhibit transitions between stable, multi-stable, periodic, and chaotic dynamics known as Hopf bifurcations~\cite{ref:neyer1982,ref:erneux2004,ref:larger2004}.

Dynamics of an optoelectronic oscillator can be described by Ikeda model~\cite{ref:ikeda1979,ref:ikeda1980,ref:erneux2004,ref:larger2004} according to which voltage $V$ at the modulation port of MZM follows the delay-differential equation
\begin{equation}
  V(t)+T_R\frac{\mathrm{d}V}{\mathrm{d}t}(t)=G^*P[V(t-\tau)],
\label{eq:ikeda}
\end{equation}
where $P[V]$ is the power transmitted by the MZM and received by the photodetector, $\tau$ is the delay time, $G^*$ is the voltage gain of the photodetector, and, $T_R$ is the response time of the system.
The transmission characteristic $P[V]$ of the Mach-Zehnder modulator is given by~\cite{ref:neyer1982,ref:erneux2004,ref:larger2004}
\begin{equation}
  P[V]=\frac{1}{2}P_\mathrm{max}\left(1+M\sin(\pi(V+V_B)/V_\pi+\phi)\right),
\label{eq:MZM-tf}
\end{equation}
where $P_\mathrm{max}$ is the total optical power in the system, $M$, $V_\pi$, and $\phi$ are the modulation depth, half-wave voltage, and intrinsic phase of the MZM, respectively, and $V_B$ is the bias voltage at the bias port of the MZM.
If the response time is much shorter than the delay time $T_R\ll\tau$, the derivative in Eq.~(\ref{eq:ikeda}) can be neglected, and we can describe the dynamics $V(t)$ by a difference equation~\cite{ref:neyer1982}
\begin{equation}
  V(t)=G^*P[V(t-\tau)],
  \label{eq:MZM-V}
\end{equation}
while analyzing $V(t)$ at discrete time steps $t=t_0+n\tau$.
Introducing the dimensionless state variable $x=V/V_\pi$, performing time multiplexing according to Eq.~(\ref{eq:multiplexing}), and adding external signal we rewrite Eq.~\ref{eq:MZM-V} in terms of a generic RC evolution Eq.~(\ref{eq:rc-evolution}) as
\begin{equation}
  \vb{x}(n+1)=\frac{G}{2}\left(1+M\sin(\beta W\vb{x}(n)+\rho W^I\vb{u}(n+1)+\Phi_0)\right),
  \label{eq:eoo-rc-evolution}
\end{equation}
where $\Phi_0=\phi+\pi V_B/V_\pi$ is the phase bias, $\beta$ and $\rho$ are the feedback and input scaling, respectively, and $G=G^*/V_\pi P_\mathrm{max}$ is the net gain.
In delay RC the matrix $W$ is generally a circulant matrix whose entries depend on the ratio of the delay time $\tau$ to the clock cycle $T$ while the matrix $W^I$ is initialized with uniformly distributed entries in the $[-1, 1]$ interval.

\subsection{FPGA-Based Delayed Feedback Tuning}
\begin{figure}[h]
  \centering
  \includegraphics[width=1\linewidth]{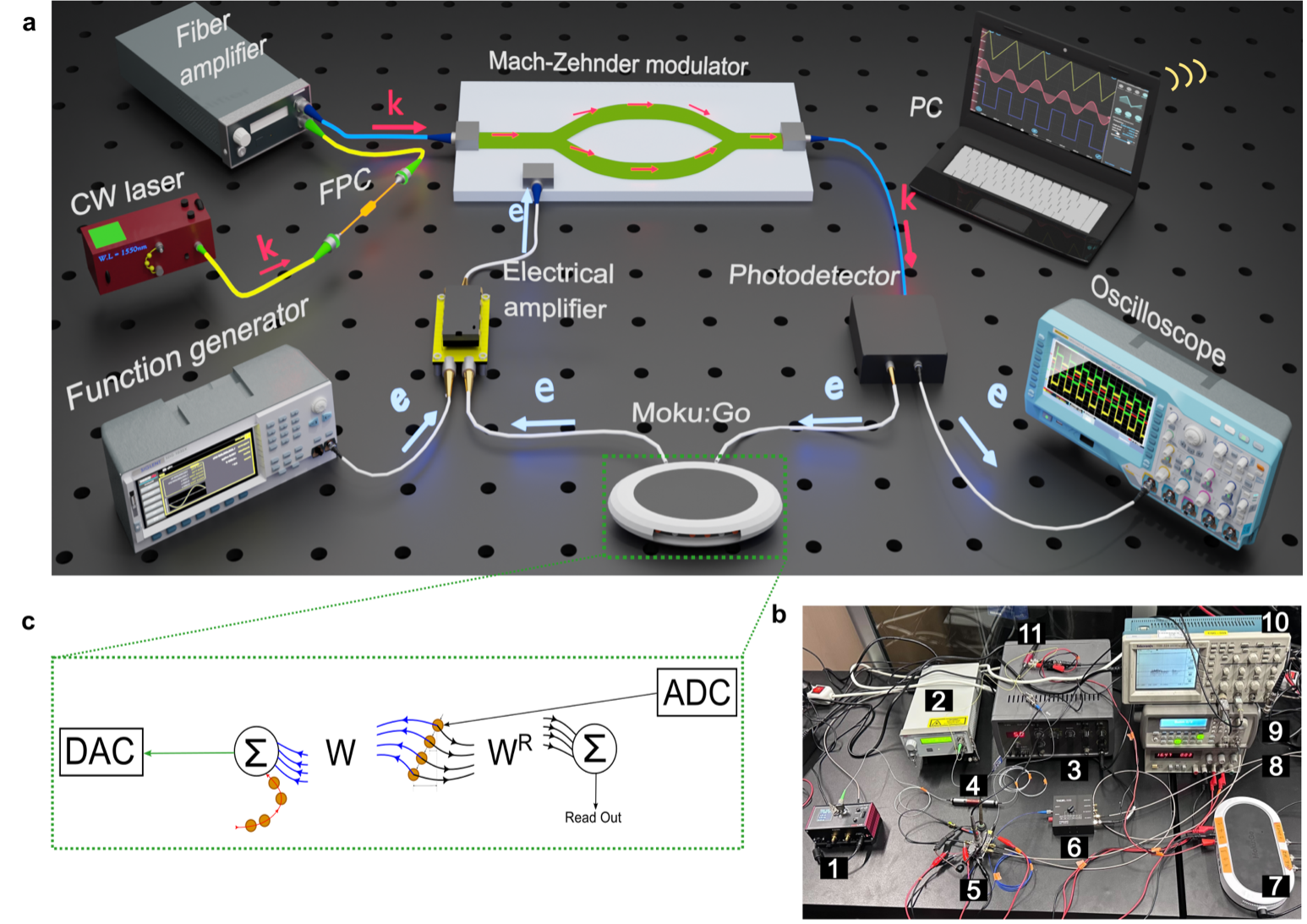}
  \caption{\textbf{a.} An artistic impression of the experimental set-up. A continuous wave (CW) laser beam is directed into the fiber polarization controller (FPC), which aligns the light polarization with the slow axis of the modulator. Subsequently, the laser beam is coupled to a fiber amplifier to maintain the system's stability. This amplified laser light is modulated with a Mach-Zehnder modulator (MZM), whose sinusoidal transmission function introduces nonlinearity into the reservoir. The modulated light is detected by a photodetector, delayed with Moku:Go's FPGA-based delay, amplified, and sent to the modulation input of the MZM, forming a closed loop. The external signal is mixed with the delayed feedback.  Note: the components are out of scale for visualization. \textbf{b.} Photograph of the experimental setup, components are labeled as follows: 1. continuous-wave (CW) laser, 2. Erbium-doped fiber amplifier (EDFA), 3. function generator producing trigger signal. 4. fiber polarization controller (FPC), 5. Mach-Zehnder modulator (MZM) with electrical driver, 6. photodetector, 7. Moku:Go, 8. power supply for the electrical driver, 9. arbitrary waveform generator (AWG), 10. oscilloscope, 11. variable optical attenuator (VOA). \textbf{c.} Schematics of the digital delay and data digitizer implemented in the Moku:Go component: delayed feedback implements the reservoir's connectivity matrix $W$ while the digitized reservoir states are weighted with the readout matrix $W^R$.}
  \label{fig:exp_setup}
\end{figure}

\FloatBarrier
The experimental setup is illustrated in Figure~\ref{fig:exp_setup}a. Figure~\ref{fig:exp_setup}b presents a photograph of the actual physical system implemented in the laboratory. Continuous-wave (CW) laser operating at 1550 nm (KLS1550, Thorlabs) was connected through a fiber polarization controller (PC1100, Fiberpro) to an Erbium-doped fiber amplifier (EDFA) (EDFA100S, Thorlabs) operating in a saturated regime.
The EDFA was used to increase the maximum optical power level, while the variable optical attenuator (VOA) was used to adjust the optical power level in the reservoir.
The attenuated optical signal was modulated with a Mach-Zehnder modulator (MZM) (LN81S-FC, Thorlabs) produced in X-cut lithium niobate.
The modulated optical signal was detected and amplified with an InGaAs photodetector (PDB450C, Thorlabs) with an embedded switchable gain trans-impedance amplifier.
The photodetector's voltage output was sent to Moku:Go (Liquid Instruments), an FPGA-based instrument that implements delay lines and data acquisition.
We note that the delayed feedback introduces the system's short-time memory property, while the MZM's sinusoidal transmission characteristic creates nonlinearity.
Moku:Go was also used as a voltage source controlling the phase bias of the MZM and attenuation of the VOA.
The output of Moku:Go was mixed with the input signal synthesized with an arbitrary waveform generator (33220A, Agilent).
The mixed signal was amplified with a driver circuit based on an operational amplifier (LM7171, Texas Instruments) to drive the MZM.

\paragraph{Delayed feedback tuning using FPGA}
Delayed feedback was implemented using an FPGA operating as the finite impulse response (FIR) filter 
\begin{equation}
  {\tilde x}(t)=\sum_{l=0}^{L-1}h_lx(t-l/r),
  \label{eq:delay-tuning}
\end{equation}
where $x(t)$ and ${\tilde x}(t)$ are the input and the output of the FIR filter, respectively, $h_l$ are the filter tap coefficients, $L$ is the filter order, and $r$ is its sampling rate.
In the experiment all FIR filter coefficients but the last were set to zero while the last coefficient was set to 1, ensuring the delay time $\tau=(L-1)/r$.
FPGA operated at either sampling rate $r$=3.906 MHz or $r$=976.6 kHz.
The filter order $L$ was varied in the range $L\in[2,232]$ at $r$=3.906 MHz and in the range $L\in[2,464]$ at $r$=976.6 kHz.
By varying FIR filter order $L$ we tuned the delay time in the range $[0.5, 60]$ and $[2, 475]$ $\mu$s at the sampling rate 3.906 MHz and 976.6 kHz, respectively.

\paragraph{Data injection and acquisition}
Reservoir transient responses were digitized using the Moku:Go's built-in datalogger operating at a sampling rate of $f=488.3$ kHz equal to the $1/8$ or $1/2$ of the sampling rate of the delayed feedback.
The time separation $\theta$ between the virtual neurons was set to $1/f$ so that each neuron corresponded to one sample in the data log.
The readout training was performed in software using routines from the reservoirpy library~\cite{ref:trouvain2020}.
To synchronize the reservoir's readout with the input signal, we employed an external function generator (label 3 in Figure~\ref{fig:exp_setup}b), producing a trigger signal that triggered bursts on the AWG and started data acquisition on the Moku:Go's datalogger.
Individual inputs were concatenated in batches to speed-up the reservoir optimization so that waveforms filled all the available AWG's memory.

\begin{figure}[htpb]
  \centering
  \includegraphics[width=0.8\linewidth]{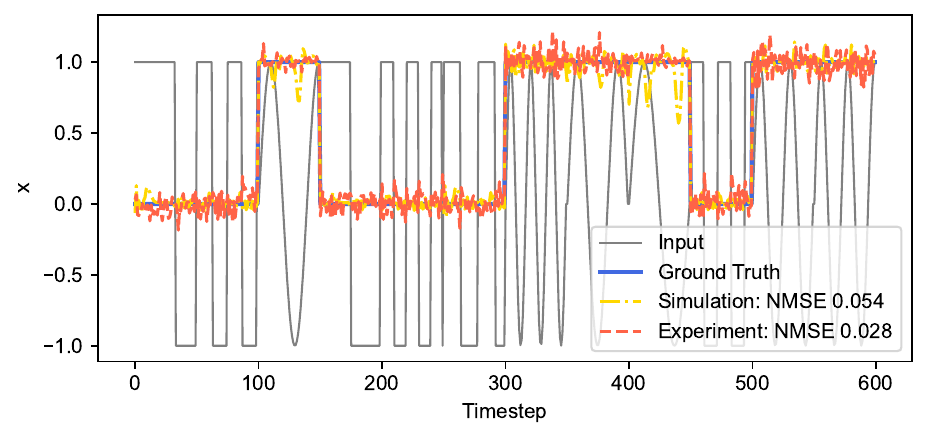}
  \caption{Sinusoidal vs. rectangular waveform classification task. The solid grey curve represents the input signal (sinusoidal and rectangular waveforms), a solid blue curve represents the target response ($x$=1 for sinusoidal, $x$=0 for rectangular waveform), the dash-dotted yellow and orange dashed curves represent the readout of simulated and \emph{in situ}-optimized experimental reservoirs, respectively. Experimental reservoir settings: input scaling $\rho$=0.19, net gain $G$=0.39, phase bias $\Phi_0$=$0.67\pi$, delay $\tau$=$0.27T$, regularization parameter $\lambda=1.4\times10^{-3}$.}
  \label{fig:sin_square}
\end{figure}

\paragraph{Sinusoidal versus rectangular waveform classification}
The \emph{in situ} optimization of our experimental RC system was performed in three benchmark tasks and compared to the simulation.
In each task five parameters were simultaneously optimized: gain $G$, phase bias $\Phi_0$, input scaling $\rho$, delay time $\tau$, and regularization parameter $\lambda$.
For the optimization, a Bayesian algorithm was employed~\cite{ref:bergstra2011,ref:trouvain2020,ref:hinaut2021} along with random search.
As a first experiment, the reservoir was trained to distinguish sinusoidal from rectangular waveforms following Paquot~\emph{et~al.}~\cite{ref:paquot2012}.
Departing from the Ref.~\cite{ref:paquot2012}. The frequency of the waveforms was varied to increase the complexity of the task.
For the training, the dataset consisting of a total of 20 waveforms was split equally into train and test sets. 
\textbf{Figure~\ref{fig:sin_square}} presents the simulated and experimental reservoir performance on the test dataset.
As the figure shows, simulated and experimental reservoirs classified waveforms.
The experimental NMSE outperformed the simulated one by almost a factor of 2: 0.028 and 0.054, respectively. Note: the simulation involves discretization errors, quantization noise, and numerical approximations that do not perfectly replicate real-world conditions.

The optimal parameter settings obtained~\emph{in situ} were: input scaling $\rho$=0.19, net gain $G$=0.39, phase bias $\Phi_0$=$0.67\pi$, delay $\tau$=$0.27T$, and regularization $\lambda=1.4\times10^{-3}$.

\FloatBarrier

\paragraph{NARMA10 time series recovery}
For a second benchmark, the reservoir was trained to predict time series generated by the Nonlinear Auto Regressive Moving Average (NARMA) model~\cite{ref:rodan2011,ref:paquot2012}, a popular benchmark task in the RC literature.

We used a NARMA model of the order ten driven by the equation
\begin{equation}
  y(n+1)=0.3y(n)+0.05y(n)\left(\sum_{i=0}^{9}y(n-i)\right)+1.5u(n-9)u(n)+0.1,
  \label{eq:narma10}
\end{equation}
analogous to the model in Refs.~\cite{ref:rodan2011,ref:paquot2012} and \cite{ref:hulser2022}.

The total length of the dataset was 8000 steps; the dataset was randomly split into train and test sets of 4000 steps length each.

\begin{figure}[htpb]
  \centering
  \includegraphics[width=0.8\linewidth]{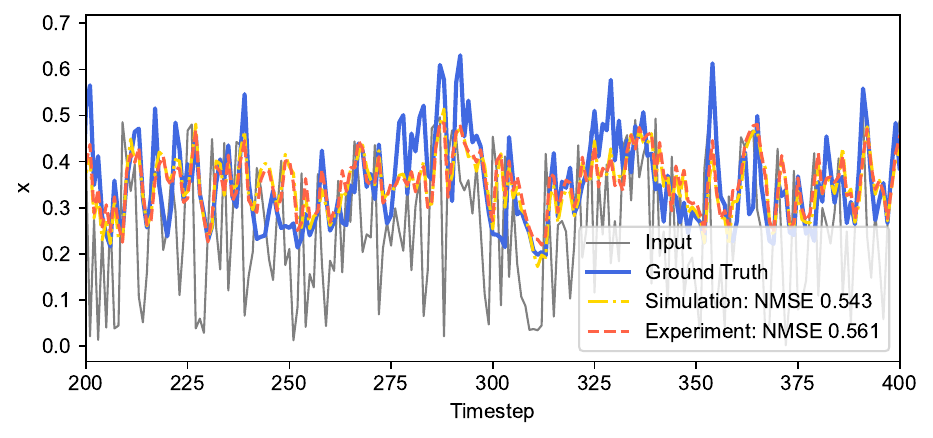}
  \caption{NARMA10 time series recovery. The solid grey curve represents the input signal (white noise), the solid blue curve represents the time series governed by the NARMA10 model \textbf{(Eq.~(\ref{eq:narma10}))}, the dash-dotted yellow and dashed orange curves represent the readout of simulated and \emph{in situ}-optimized reservoirs, respectively. Experimental reservoir settings: input scaling $\rho$=0.33, net gain $G=0.7$, phase bias $\Phi_0=0.68\pi$, delay $\tau=0.49T$, regularization parameter $\lambda=5\times10^{-3}$.}
  \label{fig:narma10}
\end{figure}

Reservoir performance in this task is presented in Figure~\ref{fig:narma10}.
One observes almost identical behavior of simulated and experimental reservoirs and similar NMSE of 0.543 vs. 0.561, respectively.

The optimal settings of the reservoir were found to be $\rho$=0.33, $G=0.7$, $\Phi_0=0.68\pi$, $\tau=0.49T$, and $\lambda=5\times10^{-3}$ for the reservoir of size 50.
The best reservoir configuration for this task performed 0.534 and 0.731 in terms of NMSE and NRMSE, respectively, at a reservoir size of 50.

NMSE in our case is considerably higher than the in Ref.~\cite{ref:paquot2012} (0.168) but is within the range of NMSEs reported in Ref.~\cite{ref:hulser2022} (0.04-0.64) for simulated delay reservoirs of the same size.
We attribute increased NMSE in our case to the larger dataset size compared to Ref.~\cite{ref:paquot2012} (8000 vs 2000 timesteps).

\FloatBarrier

\paragraph{Japanese vowels classification.}
To assess the speech recognition capability of our system, we tested it on the Japanese Vowels dataset~\cite{ref:mineichikudo1999}. 
This dataset contains a 640-time series of 12 Mel-frequency cepstrum coefficients (MFCCs) taken from recordings of nine speakers uttering a Japanese vowel.
The task is to classify the recordings by the speaker's identity.

\begin{figure}[htpb]
  \centering
  \includegraphics[width=0.9\linewidth]{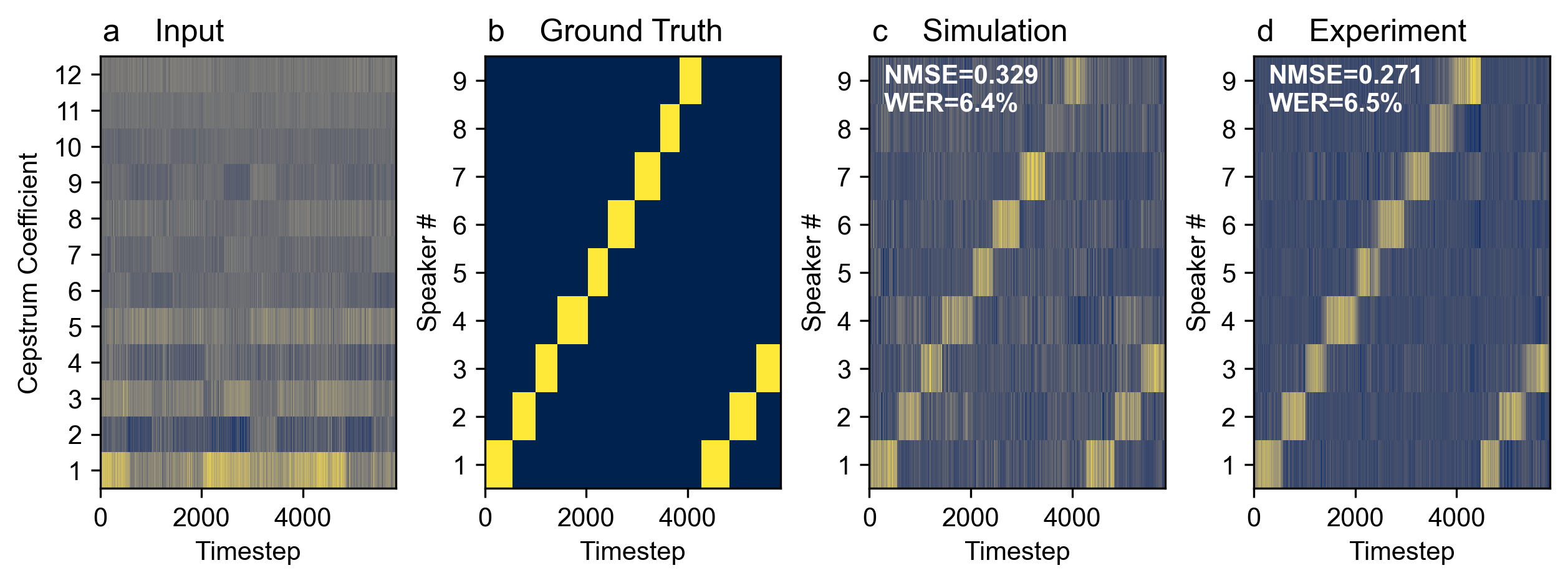}
  \caption{Japanese vowels classification task. Multiplexed waveforms of \textbf{a.} input data, \textbf{b.} ground truth, \textbf{c-d} simulated and experimental reservoirs' readouts. Reservoir settings: input scaling $\rho$=$0.47$, net gain $G$=$0.52$, phase bias $\Phi_0$=$0.44\pi$, delay $\tau$=$0.35T$, regularization parameter $\lambda=3\times10^{-7}$.}
  \label{fig:japanese_vowels}
\end{figure}

\textbf{Figure~\ref{fig:japanese_vowels}} shows the multiplexed waveforms of input data (a), target output (ground truth) (b), simulated and experimental reservoirs' readout (c-d).

In this task, \emph{in situ}-optimized reservoir outperformed simulation in terms of NMSE by about 17\% (0.271 vs 0.329) demonstrating almost identical word error rate (WER) of~6.5\% vs.~6.4\%.
Our result is comparable to the 5\% WER in Ref.~\cite{ref:picco2024} for the reservoir of size 50 and outperforms the result in~\cite{ref:paudel2020} with 18.5\% WER.

\FloatBarrier

The optimal settings of the reservoir were found to be: $\rho$=$0.47$, $G$=$0.52$, $\Phi_0$=$0.44\pi$, $\tau$=$0.35T$, and $\lambda=3\times10^{-7}$ for the reservoir with 50 nodes.

\begin{figure}
    \centering
    \includegraphics[width=\linewidth]{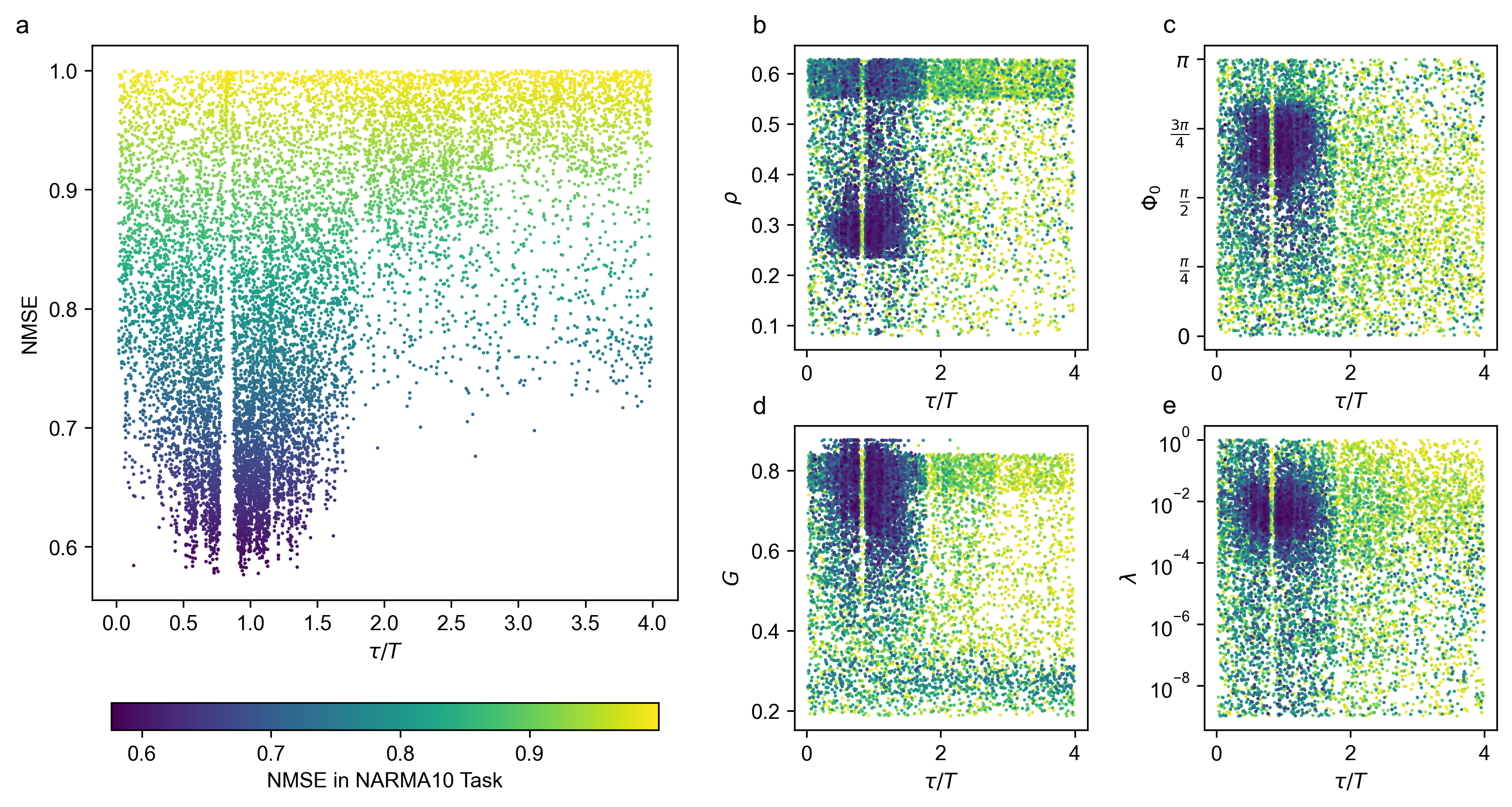}
    \caption{\emph{In situ} optimization results. Normalized mean square error (NMSE) in NARMA10 task as a function of \textbf{a}. delay to clock cycle ratio $\tau/T$ and \textbf{b}-\textbf{e} $\tau/T$ together with another parameter of the reservoir:
    \textbf{b}. $\tau/T$ and input scaling $\rho$,
    \textbf{c}. $\tau/T$ and phase bias $\Phi_0$,
    \textbf{d}. $\tau/T$ and net gain $G$,
    \textbf{e}. $\tau/T$ and regularization parameter $\lambda$.
    Each point in the plots corresponds to a tested reservoir parameter setting.}
    \label{fig:delay_effect}
\end{figure}

\paragraph{Effect of the parameters on the RC accuracy}
To illustrate the performance dependence of the system on its parameters \textbf{Figure~\ref{fig:delay_effect}} shows the results of optimization for the NARMA10 task.

We learn that for input scaling $\rho$, phase bias $\Phi_0$, and regularization parameter $\lambda$ within the studied range, there is a single local optimum centered at $\rho\approx0.3$, $\Phi_0\approx3\pi/4$, and $\lambda\approx10^{-3}$, respectively.
The picture for $G$ is more interesting: one can observe a wide local optimum above $G\approx0.6$ spanning to the top boundary and an additional local optimum near $G=0.3$.
Regarding the delay-to-clock cycle ratio, multiple narrow peaks and drops in NMSE are seen with the best performance within the range $\tau/T\in[0, 2]$.
The most prominent peak corresponds to the primary resonance $\tau/T\approx 1$ but is shifted to the left due to the latency of FPGA, resulting in underestimated $\tau$.
Noteworthy, the performance peaks and drops at specific $\tau/T$ ratios are independent of other parameters of the reservoir, confirming that the root cause for the detrimental effect on the performance are delay to clock cycle resonances that lead to the reduction of RC network complexity, as discussed in Refs.~\cite{ref:stelzer2020,ref:koster2021,ref:hulser2022}.

\FloatBarrier

\section{Discussion}
In this work, we have achieved \emph{in situ} optimization of a physical computing system based on the optoelectronic delay reservoir computing principle. 
We showed that by evaluating the system across three benchmark tasks—signal classification, time series prediction, and speech recognition—we demonstrated that \emph{in situ} optimization significantly enhances accuracy, with improvements of 17\% and 48\% in two of the tasks compared to conventional simulation-based approaches.
Beyond improving accuracy, \emph{in situ} optimization leverages the inherent speed and energy efficiency of physical computing systems while also eliminating the need for extensive numerical modeling, which often requires accounting for noise, environmental factors, and device-specific characterization.

In conclusion, reservoir computing (RC) is a powerful approach in neuromorphic computing that uses fixed, randomized internal connections to mitigate overfitting, making it highly effective for signal processing and pattern recognition tasks. Its suitability for hardware implementations across various physical platforms offers significant potential for increased computational speed and reduced energy consumption. However, achieving optimal performance in RC systems has traditionally required software-based optimization, limiting the practicality of physical computing solutions.

In this work, we introduced an \emph{in situ} optimization method for an optoelectronic delay-based RC system with digital delayed feedback. By simultaneously optimizing five key parameters, we achieved normalized mean squared error (NMSE) values of 0.028, 0.561, and 0.271 in three benchmark tasks—waveform classification, time series prediction, and speech recognition—outperforming simulation-based optimization in two of these tasks. 

This \emph{in situ} optimization method represents a significant advancement in physical computing, eliminating the need for extensive simulations and enhancing the practical applicability of RC and neuromorphic systems across real-world applications.

Overall, this study demonstrates the potential of \emph{in situ} optimization to advance the efficiency and performance of physical computing systems, offering a promising pathway for more effective implementations in real-world applications.

\section{Materials and methods}
\subsection{Mathematical Framework and Performance Evaluation of Reservoir Computing}
The evolution of the reservoir was characterized by a state vector $\vb{x}$ driven by a data signal $\vb{u}$ can be described as~\cite{ref:lukosevicius2009,ref:appeltant2011}
\begin{equation}
  \vb{x}(n+1)=f(\beta W\vb{x}(n)+\rho W^{I}\vb{u}(n+1)),
  \label{eq:rc-evolution}
\end{equation}
where $\beta$ and $\rho$ are the feedback and input scaling, respectively, $n$ is the discrete time, $f$ is a nonlinear activation function, and $W$, $W^I$ are the reservoir and input connectivity matrices, respectively.
The reservoir's transient response to the input signal was sent to a linear read-out layer where it is weighted by the readout matrix $W^R$ to obtain the output vector $\vb{y}_\mathrm{out}$ as
\begin{equation}
  \vb{y}_\mathrm{out}(n)=W^R\vb{x}(n).
  \label{eq:readout}
\end{equation}

Training of the reservoir was performed by finding the weights (elements of matrix $W^R$) minimizing the mean squared error (MSE) between the target output $\vb{y}_{t}$ and reservoir's readout given by~Eq.~(\ref{eq:readout}) $\lVert \vb{y}_{t}-W^R\vb{x}\rVert^2$ on a training dataset using linear regression.
To avoid overfitting, we minimized $\lVert \vb{y}_t-W^R\vb{x}\rVert^2+\lVert\lambda W^R\vb{x}\rVert^2$, where the second term stands for weights regularization with the regularization constant~$\lambda$ also referred to as ridge constant.
Due to the linearity of the readout layer, we found the readout weights by simple matrix inversion as
\begin{equation}
  W^R=YX^T\left(XX^T+\lambda I\right)^{-1},
\end{equation}
where $X$ and $Y$ are matrices obtained by column-wise concatenation of all reservoir states $\vb{x}(n)$ with $n=0,1,\ldots,N-1$ and all target outputs $\vb{y}_t(n)$ with $n=0,1,\ldots,N-1$, respectively, where $I$ is the identity matrix~\cite{ref:trouvain2020}.
The performance of the reservoir was evaluated by measuring the error on a validation dataset not seen during training.
We used one of the commonly used error metrics for evaluation - normalized mean squared error (NMSE), defined as
\begin{equation}
  \mathrm{NMSE}=\frac{\lVert\vb{y}-\hat{\vb{y}}\rVert^2}{\sigma^2_{\vb{y}}},
\label{eq:nmse}
\end{equation}
where $\vb{y}$ and $\hat{\vb{y}}$ are the target and the actual response on the validation dataset, respectively, while $\sigma^2_{\vb{y}}$ is the variance of the target response~\cite{ref:botchkarev2019}.
A related metric is the normalized root mean square error (NRMSE), which is the square root of NMSE.

\subsection{Mechanism of Delay-Based Reservoir Computing}
To relate temporal dynamics of a delay-based reservoir with the evolution equation~(\ref{eq:rc-evolution}) the values of the state variable $x(t)$ sampled at evenly spaced instants of time during a clock cycle period $T$ are mapped via time multiplexing to the components of the reservoir's state vector $\vb{x}(n)\equiv x^i(n)$ forming a set of virtual neurons as shown in Figure~\ref{fig:rc_scheme}b
\begin{equation}
    \vb{x}(n)\equiv x^{i}(n)=x(i\theta+nT),
\label{eq:multiplexing}
\end{equation}
where $i=0,\ldots,k-1$, $\theta=T/k$ is the virtual neurons temporal separation with $k$ being the size of the reservoir.
To map the input signal onto the virtual nodes of the reservoir, input masking is performed by applying sample-and-hold operation onto the input and multiplying by a periodic piecewise-constant function with the period $T$~\cite{ref:appeltant2011,ref:brunner2018}.
For establishing interactions between different neurons in successive layers (expressed by off-diagonal entries in the matrix $W$ of Eq.~(\ref{eq:rc-evolution})) several approaches to time multiplexing exist: in~\cite{ref:larger2012} delay and clock cycle were synchronous $\tau=T$ while an introduced low-pass transient characteristic caused neighboring neurons to interact, in~\cite{ref:martinenghi2012}, analogously, delay and clock cycle were synchronized, but multiple fractional delays were introduced.
It was observed by Rodan~\emph{et al.}~\cite{ref:rodan2011} and first exploited in hardware by Paquot~\emph{et al.}~\cite{ref:paquot2012} that desynchronization of the clock cycle and delay time leads to a more interconnected reservoir topology by causing different neurons in the successive recurrent layers to interact. 
Specifically, in~\cite{ref:paquot2012,ref:antonik2017} delay time was set to $\tau=T+\theta$ providing minimum complexity network structure suggested in~\cite{ref:rodan2011}.
The structure of a delay-based RC is schematically shown in Figure~\ref{fig:rc_scheme}b.

\medskip
\textbf{Acknowledgements} \par 
The research was supported by the EU ERA-NET DIEGO project Ministry of Energy, Grant No. 221-11-032, and the Lower Saxony's Minister of Science and Culture and the Volkswagen Foundation under the program "Zukunft.niedersachsen: Research Cooperation Lower Saxony – Israel" 76251-5615/2023.

\medskip
\textbf{Author contributions} \par
Fyodor Morozko and Shadad Watad contributed equally to the conceptualization, methodology, investigation, and formal analysis. Fyodor Morozko developed software. All the authors wrote the original draft of the manuscript. Fyodor Morozko, Shadad Watad, Amir Naser, and Alina Karabchevsky reviewed and edited the manuscript. Alina Karabchevsky supervised the project and acquired funding.

\medskip
\textbf{Conflict of interest} \par
The authors declare that they have no competing interests.

\medskip

\bibliographystyle{MSP}
\bibliography{bibliography}


\FloatBarrier

\begin{figure}
\textbf{Table of Contents}\\
\medskip
  \includegraphics[width=0.6\linewidth]{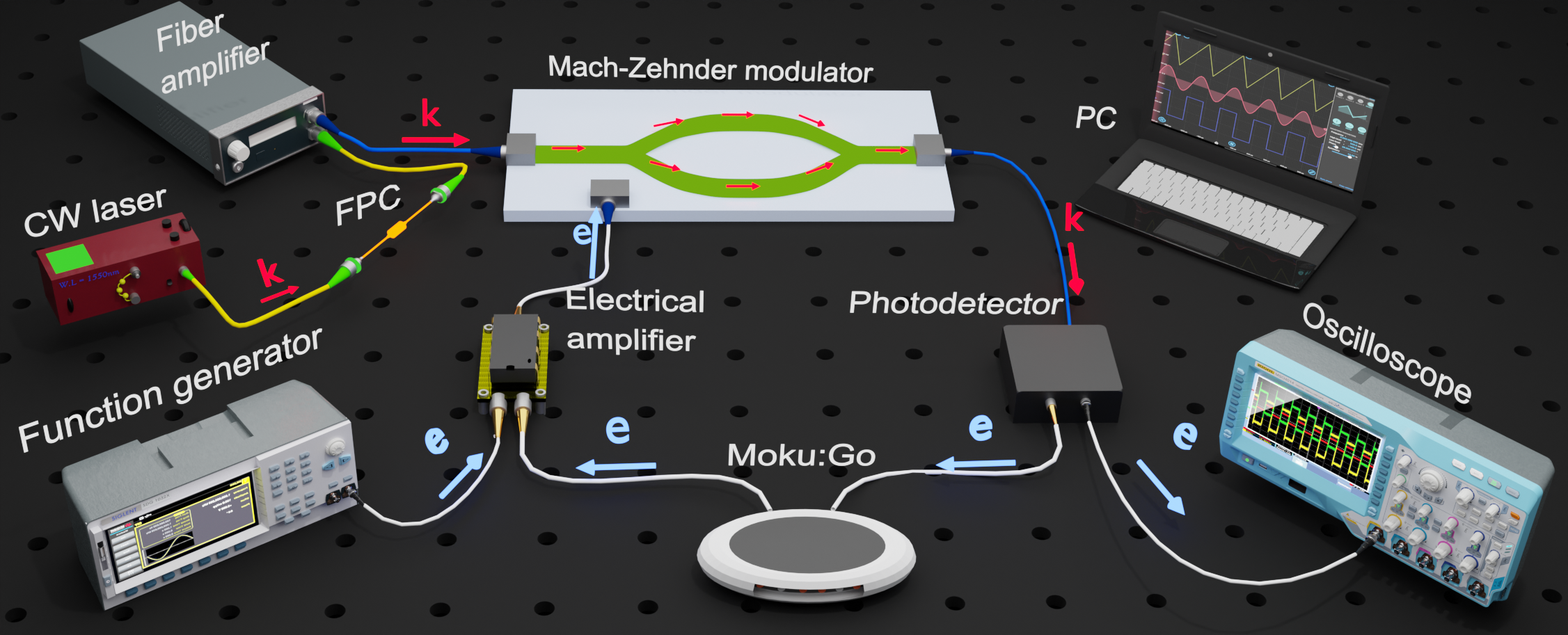}
  \medskip
  \caption*{
    In-situ optimization of multiple parameters of an optoelectronic delay-based reservoir computing system is realized and tested in three benchmark tasks.
    Accuracy achieved with \emph{in situ} optimization exceeded accuracy obtained with simulation by 17\% and 48\% in two of the three tasks.
    Experimental results reveal that the delay timescale is critical for the accuracy of delay-based reservoir computing.
  }
\end{figure}

\end{document}


\title{In Situ Optimization of an Optoelectronic Reservoir Computer with Digital Delayed Feedback}

\maketitle


\section{Dynamics of the optoelectronic oscillator}
\begin{figure}[h]
  \includegraphics[width=\linewidth]{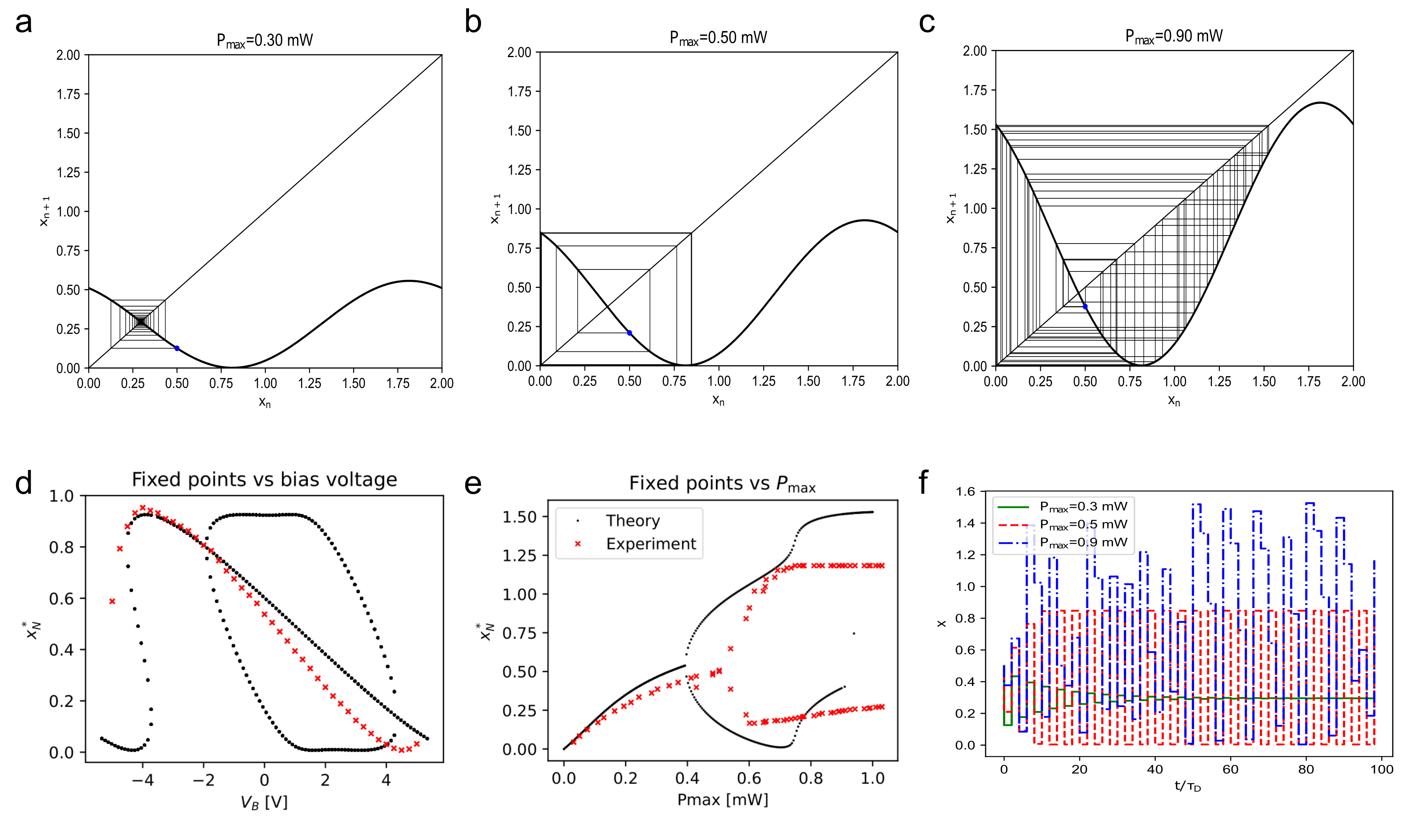}
    \caption{Simulated and experimental results from set-up shown in Figure 2 in the main text: calculated results based on the discrete-time model: (a) stable, (b) periodic, and (c) chaotic regimes. (d-e) Fixed points of the system as a function of (d) bias voltage $V_B$ and (e) optical power $P_\mathrm{max}$, f) Simulated dynamics using discrete-time model
    }
  \label{fig:dynamics}
\end{figure}

\textbf{Figure \ref{fig:dynamics}} presents the behavior of the optoelectronic oscillator: Figures (a-c) show cobweb diagrams in stable, periodic, and chaotic regimes, respectively, Figures d-e show simulated and experimentally obtained fixed points in the systems at different settings based on the Ikeda model~\cite{ref:ikeda1979,ref:ikeda1980,ref:ikeda1982,ref:neyer1982,ref:larger2004}.

As discussed in~\cite{ref:neyer1982}, if the response time of the system $\tau$ is much less than the delay time $\tau_{D}$ $\tau\ll\tau_D$ the continuous-time dynamics of the delayed-feedback system can be efficiently modeled by the discrete-time difference equation
\begin{equation}
  x_{n+1}=G/2\left(1+M\sin(\pi(x_n+x_b))\right),
  \label{eq:ikeda-discrete}
\end{equation}
where $x_{n}=V(n\tau_D)/V_\pi$, $x_b=V_B/V_\pi$, and
\begin{equation}
  G=P_\mathrm{out}G^{*}/V_\pi
\end{equation}
is the net gain of the open loop. The graphical solution of the \ref{eq:ikeda-discrete} is depicted in \textbf{Figure }. The resulting path originates from $x_\mathrm{0}$ in the proximity of $x_\mathrm{1}$ and moves away from $x_\mathrm{1}$ due to instability. However, it is eventually attracted to a stable limit determined by the number of periods. For \textbf{Figure \ref{fig:dynamics}a}, the period is 1, for \textbf{Figure \ref{fig:dynamics}b} it is 2, and for \textbf{Figure \ref{fig:dynamics}c}, the period is 3.

The simulation results of the optoelectronic oscillator are represented in subplots \textbf{Figures \ref{fig:dynamics}a-c}, which show the cobweb diagrams of the different regimes.
The method involves overlapping the cobweb plot with the function $y=x$ to identify the fixed points.
\textbf{Figure \ref{fig:dynamics}a} shows the dynamics of the system in the stable regime corresponding to the parameters $P_\mathrm{max}=0.3$ mW denotes the maximum power, $V_\mathrm{b}=0$ V represents the bias voltage, $G=0.56$ signifies the feedback gain, and $M=0.983$ is defined as the modulation factor.
So, in the intersection, we obtain one unstable fixed point.
\textbf{Figure \ref{fig:dynamics}b} shows the periodic regime with $P_\mathrm{max}=0.5$ mW, $V_\mathrm{b}=0$ V and $G=0.93$. Here, we encounter two unstable fixed points.
The chaotic regime is shown in  \textbf{Figure \ref{fig:dynamics}c} with $P_\mathrm{max}=0.9$ mW and \textbf{$V_\mathrm{b}=0$ V}, $G=1.49$.
By examining the three subplots, we can observe that the stability of $x_\mathrm{n+1}$ vs. $x_\mathrm{n}$ is decreasing compared to the stable regime in \textbf{Figure \ref{fig:dynamics}a}.
Hopf bifurcations are shown in \textbf{Figures \ref{fig:dynamics}d-f}. Where  \textbf{Figures \ref{fig:dynamics}d} demonstrates the equilibrium values of the first iteration "N=1" for the "Number of the equation" considering the bias voltage. The black curve represents the simulated results, while the red curve represents the experimental measurements, considering the parameters \textbf{{$G=0.93$}} and \textbf{$M=0.983$}. The graph showcases stable regions with a single stable solution denoting system stability. In contrast, the bistable and periodic regions exhibit three solutions: two stable and one unstable. The presence of periodic solutions arises from the equation's bifurcation of stable states. Simulated results are used to validate the theoretical predictions, ensuring the accuracy and reliability of the findings.\textbf{Figures \ref{fig:dynamics}e} which exhibits a cascade of periodic solutions with fixed parameters \textbf{$G=0.93$} and \textbf{$M=0.983$}. The graph presents a depiction of the stable and unstable fixed points as a function of the maximum power, $P_\mathrm{max}$, up to iteration \textbf{$N=8$}. A significant observation is that as $P_\mathrm{max}$ increases, so does the net feedback gain, G, leading to a growing number of bifurcations. Notably, this cascade of periodic solutions appears to extend infinitely until reaching the critical value \textbf{$G_{c}$=1.49}. Upon reaching it, the system transforms into an aperiodic(chaotic) regime, resulting in the absence of periodic solutions.
\textbf{Figure \ref{fig:dynamics}f} depicts stable, periodic, and chaotic regimes in the time domain.


\section{Electro-optic modulator driver}

\begin{figure}[h]
  \centering
  \includegraphics[width=0.8\linewidth]{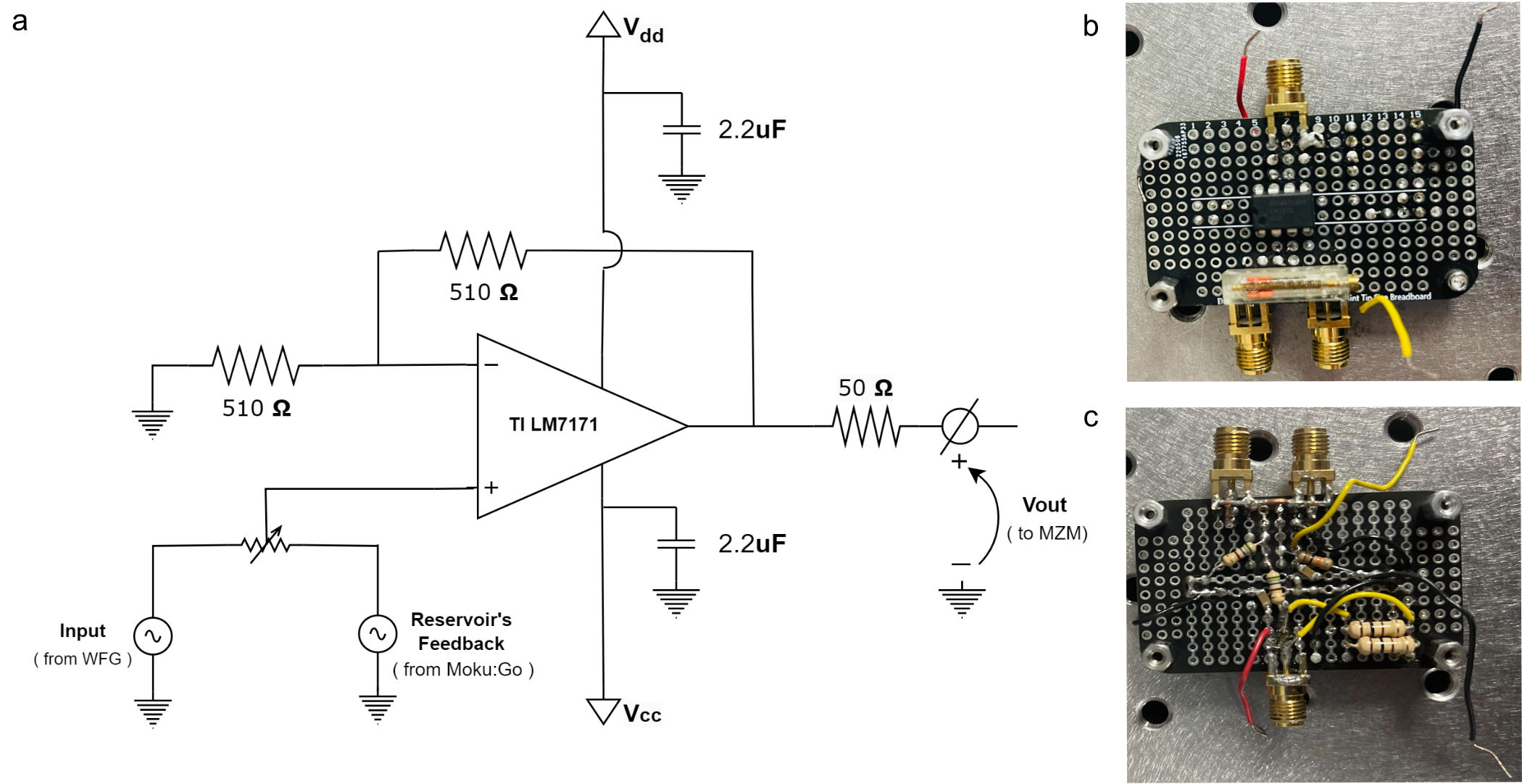}
    \caption{(a) Schematic of the electro-optic modulator driver. (b-c) design of the amplifier board.}
  \label{fig:schematic}
\end{figure}

To drive the electro-optical modulator which has $50~\Omega$ characteristic impedance the low-current output signal of the Moku:Go's FIR filter needed to be amplified.
For this purpose we have built a driver circuit based on a high-speed analog operational amplifier (LM7171, Texas Instruments) as shown in Figure~\ref{fig:schematic}.
We have incorporated a voltage divider at the input of the amplifier to adjust the relative strengths of the delayed feedback and input signal.

\FloatBarrier

\bibliographystyle{plain}
\bibliography{bibliography}